\documentstyle[prl,aps,epsf,multicol]{revtex}

\begin{document}
\draft

\title{Inelastic Neutron Scattering Signal from Deconfined Spinons in a
Fractionalized Antiferromagnet}    
\author{C. Lannert$^1$ and Matthew P. A. Fisher$^2$}
\address{$^1$Department of Physics, University of California, Santa
Barbara, CA 93106 \\ 
$^2$Institute for Theoretical Physics, University of California, 
Santa Barbara, CA 93106--4030}  

\date{\today}
\maketitle

\begin{abstract}
We calculate the contribution of deconfined spinons to inelastic neutron
scattering (INS) in 
the fractionalized antiferromagnet ($AF^*$), introduced elsewhere. We
find 
that the presence of free spin-1/2 charge-less
excitations leads to a continuum INS signal above the
N\'{e}el gap. This signal is found above and in addition to the usual
spin-1 magnon signal, which to lowest order is the same as in
the more conventional confined antiferromagnet. We calculate the
relative weights of these two signals and find that the spinons
contribute to the longitudinal response, where the magnon signal is
absent to lowest order. Possible higher-order effects of interactions 
between magnons and spinons in the $AF^*$ phase are also discussed. 
\end{abstract}
\vspace{0.15cm}

\begin{multicols}{2}
\narrowtext 

\section{introduction}

Theories of spin-charge separation in the high-$T_c$ cuprates
have been hotly debated almost since the original discovery of these
materials \cite{forefathers}.  
Finding a theory of electrons in more than one spatial dimension which
exhibits zero-temperature spin-charge separation has proved to be as
theoretically challenging as it is phenomenologically appealing
\cite{usualsuspects}. It
can be argued that all such theories will admit, in the low-energy
limit, a formulation in terms of a $Z_2$ gauge theory
\cite{U1Z2}. Recent papers have addressed the problem of finding
microscopic models of electrons which become fractionalized in some
range of their parameters \cite{bfg&s}. 
It remains an important task to enumerate concrete, experimentally-
measurable consequences of these exciting theoretical
ideas. Previously, we have explored the consequences of
two-dimensional fractionalization on 
the spectral function, as probed by angle-resolved photo-emission
spectroscopy \cite{arpes}. In this
paper, we calculate the inelastic neutron scattering signal from 
spinons in a 
fractionalized antiferromagnet ($AF^*$). We find that these spin-1/2,
charge-less excitations lead to a continuum
of excitations above a gap. Because we are interested
in the parent insulators of cuprate superconductors, we have taken a
phenomenological model for the spinons which gives them both a N\'{e}el
gap arising from antiferromagnetic ordering and a \emph{d}-wave
pairing gap which becomes 
the pseudogap at moderate doping and the superconducting gap in
the superconducting phase. We contrast this signal with the signal
from excitations in a conventional antiferromagnet and calculate the
strength of the spinon signal compared to the magnon signal (which is
also present). This comparison estimates the feasibility of measuring this 
anomalous signal in the parent insulators. We also discuss 
higher-order effects stemming from interactions between spinons and
magnons.  

\section{the model \label{model}} 

The $AF^*$ phase has been discussed elsewhere \cite{Z2,NL,CC}
and here we use the same phenomenological model introduced and
justified \cite{arpes,Z2} previously. We assume that the steps
of: (1) deriving a lattice Hamiltonian containing all
important effective
interactions between electrons and (2) splitting the electron into
chargon 
and spinon fields ($c_{i\alpha} = b_i f_{i\alpha}$) and deriving the
appropriate $Z_2$ gauge theory have 
been performed and we have arrived at the following effective
low-energy Hamiltonian for the system in 2-dimensional fractionalized
phases: 
\begin{eqnarray}
H &=& \sum_{<ij>} [-t_s \hat{f}_{i\alpha}^{\dag}\hat{f}_{j\alpha}
+ \Delta_{ij}\hat{f}_{i\uparrow}\hat{f}_{j\downarrow} - t_c
\hat{b}_i^{\dag}\hat{b}_j + H.c.]  \nonumber \\
& & + U \sum_i [\hat{b}^{\dag}_i\hat{b}_i -( 1-x)]^2 + H_g,
\label{separated} \\
H_g &=& g\sum_{<i,j>} \hat{\mathbf{S}}_i \cdot
\hat{\mathbf{S}}_j, 
\label{Hg} 
\end{eqnarray}
where the spinon pairing $\Delta_{ij}$ is taken to be
\emph{d}-wave:
\begin{equation}
\Delta_{ij} = \left\{ \begin{array}{ll} +\Delta &
\mbox{along} \;\; \hat{x}, \\ -\Delta & \mbox{along} \;\; \hat{y}, \\
\end{array} \right. 
\end{equation}
and the spin operator is $\hat{\mathbf{S}}_i = \frac{1}{2}
\hat{f}^{\dag}_i \mbox{\boldmath $\sigma$} \hat{f}_i $. Here,
$\scriptstyle{<i,j>}$ are nearest neighbors on a 2d square lattice.
The $U$ term is a Hubbard-like interaction for $(1-x)$ chargons per
unit cell. 

\begin{figure}
\epsfxsize=3.5in
\centerline{\epsffile{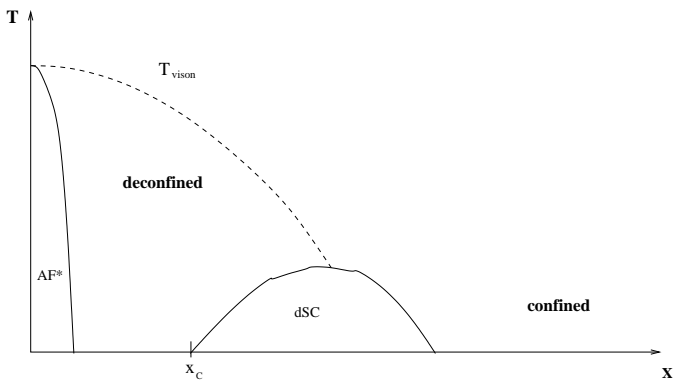}}
\vspace{0.15in}
\caption{Schematic phase diagram for the high $T_c$ cuprates within a
spin-charge separation scenario.}
\vspace{0.15in}
\label{phasediagram}
\end{figure}

We now briefly justify this model for the underdoped cuprate materials on
phenomenological grounds.  
For sufficiently small doping and low temperatures such that the
$Z_2$ theory exhibits fractionalization, the Hamiltonian is as written
in Eq.(\ref{separated}). At temperatures below the energy scale
$\Delta$, the spinons are effectively paired into \emph{d}-wave
singlets and there is a \emph{d}-wave gap to spin-1/2
excitations. For large enough $g$ (and an additional minuscule 3d spin
coupling) the 
system develops long-range antiferromagnetic order.   
At half-filling, the chargons are gapped into
an insulating phase and we obtain a
fractionalized insulator with long-range N\'{e}el order and an
additional \emph{d}-wave gap to spin-1/2 excitations, previously
dubbed $AF^*$ \cite{Z2}. Moving away  
from half-filling, the antiferromagnetic order will be quickly
suppressed, while for $t_c \ll U$ and with an additional long-range
Coulomb 
interaction, one still expects the chargons to be insulating. We then
have a fractionalized insulating phase with a \emph{d}-wave gap to
spin-1/2 excitations. Within a spin-charge separation scenario, this
phase is identified with the 
pseudogap regime in the cuprates. For chargon hopping, $t_c$,
sufficiently large, the 
chargons Bose condense, giving a \emph{d}-wave superconductor. At
large dopings, we expect the system to recover Fermi liquid
properties, as occurs when the vortex excitations (visons) in the
Ising gauge field condense thereby confining the spinons and chargons
to form the electron. A schematic phase diagram is shown in
Fig. \ref{phasediagram}. 
In this paper, we elucidate further some of the properties of the
$AF^*$ phase, found at half-filling. 

Recent experiments by Bonn, Moler,{\em  et al} put limits on the
likelihood of this sort of spin-charge separation in
$Y{Ba}_{2}{Cu}_{3}{O}_{6+x}$, 
although the experiments have only been performed on one sample
so far \cite{MB}. The question of spin response in an antiferromagnet
which is fractionalized is nevertheless well-posed and could be
relevant to  other materials. Also, it is quite possible that some
other sort of exotic order lurks in the cuprates; this work would then
serve as an illustrative calculation.

\section{Effective Hamiltonian for the Spin Sector}

In this paper we work at half-filling, where the charge
degrees of freedom will be gapped into a Mott insulating phase, and
calculate the spin response of the system, appropriate for magnetic
probes such as neutron scattering. Hence, from here on, we assume that
the relevant piece of the Hamiltonian in Eq.(\ref{separated}) is that
containing the spin degrees of freedom and we ignore the charge
degrees of freedom. At temperatures much less than the vison energy,
the chargons and spinons are essentially non-interacting, so this is
reasonable in a fractionalized phase.

\begin{figure}
\epsfxsize=2.5in
\centerline{\epsffile{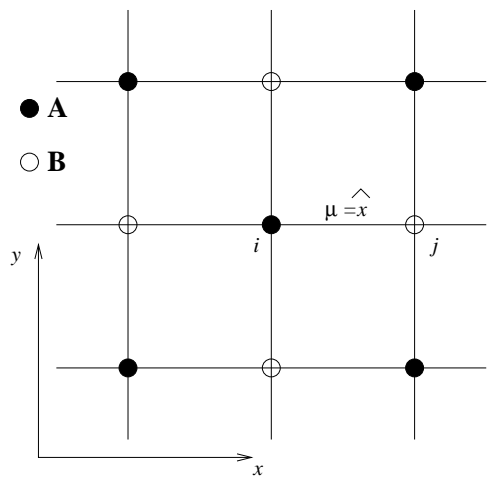}}
\vspace{0.15in}
\caption{The 2d square lattice with sublattices A and B marked.}
\vspace{0.15in}
\label{sublattices}
\end{figure}

$H_g$ (Eq.(\ref{Hg})) may be decoupled in a path integral, using a
Hubbard-Stratonovich transformation. This gives us the following
low-energy theory for the spin sector: 
\begin{eqnarray}
\lefteqn{H_{\mbox{\scriptsize \em spin}} = \sum_{<i,j>} [-t_s
\hat{f}^{\dag}_i\hat{f}_j + \Delta_{ij} \hat{f}_{i \uparrow}
\hat{f}_{j \downarrow} +H.c.]} \nonumber \\
& &  -g\sum_{i \in A, \mu} {\mathbf{N}}_{i, \mu} \cdot \left(
\hat{\mathbf{S}}_i - \hat{\mathbf{S}}_{i+\mu}
\right) + \frac{g}{2} \sum_{i \in A, \mu} {\left(
{\mathbf{N}}_{i,\mu}\right) }^2,  
\label{wholeshebang}
\end{eqnarray}
where $\mathbf{N}$ is a 3-component vector of classical fields living
on the nearest-neighbor links of the lattice, which we have broken
into its two square sublattices, labeled $A$ and $B$ and shown in
Fig. \ref{sublattices}. $\mu \in \{ \pm \hat{x}, \pm \hat{y} \} $. 

We begin by analyzing Eq.(\ref{wholeshebang}) at the
mean field level. We find antiferromagnetic N\'{e}el
ordering and a quadratic theory for the spinons which can be
solved exactly. We derive a self-consistency equation for the mean
field approximation which connects the magnitude of the N\'{e}el order to
properties of the spinons. Fluctuations about this mean field solution
lead to spin-waves (as in the conventional antiferromagnet). 
Thus, we find that the fractionalized antiferromagnetic phase has two
spin-carrying excitations: the spin-1 magnons and the spin-1/2 spinons,
which interact with each other. To deal with these interactions, we
treat the spinons as a perturbation: setting $t_s = \Delta = 0$, one
recovers the conventional antiferromagnet and its spin-wave
excitations. Integrating out
the spinons to each order in $t_s/g$ and $\Delta/g$ would give
modifications of the spin-wave theory due to the spinons. If one
wishes to find higher-order properties of the spinons, one may
start with the mean field result and then integrate out the
magnons, generating interactions between the spinons. 
In the limit $t_s,\Delta \ll g$, it is clear that the theory in
Eq.(\ref{wholeshebang}) can be solved in a controlled fashion.

\subsection{Mean Field Theory \label{mft}}

First, we ignore the spinons, setting $t_s
= \Delta = 0$ and reducing 
$\hat{\mathbf{S}}_i$ to the usual spin-1/2 quantum
operator. Eq.(\ref{wholeshebang}) becomes:
\begin{eqnarray}
H_g^{\mbox{\scriptsize \em eff}} &=& \frac{g}{2} \sum_{i \in A}
\sum_{\mu} {\mathbf{N}}_{i,\mu}  
\cdot {\mathbf{N}}_{i, \mu} \nonumber \\ 
 & & - g \sum_{i \in A} \sum_{\mu} {\mathbf{N}}_{i,\mu} \cdot
 \left(\hat{\mathbf{S}}_i - \hat{\mathbf{S}}_{i+\mu} \right) .  
\label{HS} 
\end{eqnarray} 
Choosing $\hat{z}$ as the spin quantization axis, this is
minimized classically by: 
\begin{eqnarray}
\hat{\mathrm{S}}_i^z |\uparrow \rangle _{i \in A} &=& \frac{1}{2} |
\uparrow \rangle_{i \in A}, \\
\hat{\mathrm{S}}^z_j | \downarrow \rangle_{j \in B} &=& -\frac{1}{2}
|\downarrow \rangle_{j \in B}, 
\end{eqnarray}
leading to a Hamiltonian for the field $\mathbf{N}$:
\begin{equation}
H_g^{\mbox{\scriptsize \em eff}} = \frac{g}{2} \sum_{i\in A}
\sum_{\mu} {\left( {\mathbf{N}}_{i,\mu} \right) }^2 -
g\sum_{i\in A} \sum_{\mu}{\mathrm{N}}^z_{i,\mu}.  
\end{equation}
This is minimized by: $\langle {\mathrm{N}}^z_{i,\mu} \rangle \equiv
N_0 = 1 \; \forall \, i, \mu$. 

Plugging this mean field solution for $\mathbf{N}$ back into 
Eq.(\ref{wholeshebang}) gives a Hamiltonian for the spin sector at the
mean field level which is quadratic in the spinons:
\begin{eqnarray}
H_{\mbox
{\scriptsize \em spin}}^{\mbox{\scriptsize MFT}} &=& \sum_{<i,j>} [-t_s
\hat{f}^{\dag}_{i \alpha} \hat{f}_{j \alpha} + 
\Delta_{ij} \hat{f}_{i \uparrow} \hat{f}_{j \downarrow} + H.c.] \nonumber \\
& & - 4g \sum_i N_0 \hat{z} 
\cdot \hat{\mathbf{S}}_i^{\mbox{\boldmath $\scriptstyle \pi$}} +
\frac{g}{2}\sum_{i \in A,\mu} N_0^2,  
\label{MFT}
\end{eqnarray}
where we have retained $\hat{z}$ as the spin quantization axis and
have written $\sum_i (-1)^{x+y} \hat{\mathbf{S}}_i \equiv \sum_i
\hat{\mathbf{S}}^{\mbox{\boldmath $\scriptscriptstyle \pi$}}_i =
\frac{1}{2} \sum_{\mathbf{k}} 
\hat{f}^{\dag}_{\mathbf{k} + \mbox{\boldmath $\scriptscriptstyle
\pi$}} \mbox{\boldmath $\sigma$}
\hat{f}_{\mathbf{k}}$.  
The lattice spacing has been set to unity. 

The full solution to this quadratic spinon Hamiltonian has been given
elsewhere \cite{arpes,gros} and here we reproduce only the dispersion:
\begin{equation}
{\cal E}_{\mathbf{k}}^2 = N_g^2 + \epsilon_{\mathbf{k}}^2 + \Delta_{\mathbf{k}}^2 ,
\label{fdispersioni}
\end{equation}
with
\begin{eqnarray}
N_g &=& 2gN_0 , \\
\epsilon_{\mathbf{k}} &=& -2 t_s (\cos k_x + \cos k_y), \\
\Delta_{\mathbf{k}} &=& -\Delta (\cos k_x - \cos k_y).
\label{fdispersionf}
\end{eqnarray}
  
\subsection{Self-Consistency of the Mean Field Solution \label{scmft}}

The mean field solution with $N_0 = 1$ is found formally in the limit
$t_s = \Delta =0$, and one expects the spinons to reduce the N\'{e}el
order 
from this maximum value. We therefore take a mean field solution
of the form $\langle {\mathrm{N}}^z_{i,\mu} \rangle = N_0 $ and demand
that it be a saddle point of the full theory, Eq.(\ref{MFT}):
\begin{eqnarray}
Z[N_0] &=& \mbox{Tr}_{\{\hat{f},\hat{f}^{\dag}\}} e^{-\beta
H^{\mbox{\tiny MFT}}[N_0,\hat{f},\hat{f}^{\dag}]}, \\ 
Z[N_0] &\equiv & e^{-S^{\mbox{\tiny MF}}[N_0]}. 
\end{eqnarray}
Tracing out the quadratic spinons gives an expression for
$S^{\mbox{\tiny MF}}[N_0]$ and the saddle point condition,
$\delta S^{\mbox{\tiny MF}} / \delta N_0 = 0$, gives a
self-consistent equation 
for $N_0$. At zero temperature, this is:
\begin{equation}
1 = 2g \int \frac{d^2k}{(2\pi)^2} \; \frac{1}{{\cal E}_k[N_0]},
\label{selfcon}
\end{equation}
with ${\cal E}_k[N_0]$ given in Eq.(\ref{fdispersioni}).

It is easy to check that for $t_s = \Delta = 0$, Eq.(\ref{selfcon})
gives $N_0 =1$. For $t_s, \Delta \ll g$, one finds the perturbative
result: 
\begin{equation}
N_0 \simeq 1 - \frac{(2t_s)^2 + \Delta^2}{8 g^2} + \cdots .
\label{expansion} 
\end{equation}
At large values of $t_s/g$ and $\Delta/g$, one expects the
spinons to drive the N\'{e}el order to zero, even at zero temperature.
In order to keep our calculations controlled, we work in the
limit $t_s, \Delta \ll g$ and treat the spinons as a perturbation. The
reasonableness of this limit for  
the physical systems in question will be discussed later.

\subsection{Fluctuations About the Mean Field \label{fluctuations}}

We see from the self-consistent mean field calculation of the previous
section that the
N\'{e}el order is reduced at zero temperature by the spinons. Even in
the absence of the spinons (i.e. in the pure-spin model with $t_s =
\Delta = 0$) we know that fluctuations in the order parameter are
important and lead to magnons. This suggests the following program for
calculating the spin excitations in $AF^*$ beyond the mean field
level. First, set $t_s = \Delta =0$ and work with the spin-1/2 quantum
operator, $\hat{\mathbf{S}}_i$. ``Integrating out'' these operators on 
each site leads to an effective theory of fluctuations in the
field $\mathbf{N}$ and gives the spin-wave dispersion. 
Then, one may integrate out the spinons
perturbatively in $t_s/g$ and $\Delta/g$. This will lead to 
interactions between the magnons and give corrections to the
magnon dispersion.

With $t_s$ and $\Delta$ set to zero, the effective spin Hamiltonian,
Eq.(\ref{wholeshebang}), reduces to $H_g^{\mbox{\tiny \em
eff}}[{\mathbf{N}}, \hat{\mathbf{S}}]$,
Eq.(\ref{HS}). In this section we look 
at fluctuations of the field $\mathbf{N}$ around the mean field
solution. 
We therefore set ${\mathrm{N}}^z=1$ on all links and look at the
fluctuations, $ {\mathbf{N}} - \hat{z} {\mathrm{N}}^z \simeq
{\mathbf{N}}^{\bot}$. $H_g^{\mbox{\tiny \em eff}}$ can then be written
in the following form: 
\begin{eqnarray}
& &H^{\mbox{\scriptsize \em eff}}_g = \frac{g}{2}\sum_{i \in
A}\sum_{\mu} {\left( {\mathbf{N}}^{\bot}_{i,\mu}\right) }^2 +
H^{\mbox{\scriptsize int}}, \\ 
& &H^{\mbox{\scriptsize int}} = H_0 + H_1 , \label{spqm1} \\
& &H_0 = -4g\sum_{i \in A} \hat{\mathrm{S}}^z_i  + 4g\sum_{j \in B}
\hat{\mathrm{S}}^z_j ,\\ 
& &H_1 = -g\sum_{i \in A , \mu} \hat{\mathbf{S}}^{\bot}_i
\cdot {\mathbf{N}}_{i,\mu}^{\bot} + g \sum_{j \in B ,\mu} 
\hat{\mathbf{S}}^{\bot}_j \cdot {\mathbf{N}}_{j-\mu,\mu}^{\bot} . 
\label{spqm2}   
\end{eqnarray} 
Integrating out the operators $\hat{\mathbf{S}}_i$ on each site
amounts to 
performing perturbation theory in $H^{\mbox{\tiny int}}$. To second order,
the resulting effective action for ${\mathbf{N}}^{\bot}$ is: 
\begin{eqnarray}
\lefteqn{exp\{-S^{\mbox{\tiny \em eff}}[{\mathbf{N}}^{\bot}]\} =
\mbox{Tr}_{\{\hat{\mathbf{S}}\}} e^{-\beta H_g^{\mbox{\tiny
\em eff}}[{\mathbf{N}}, \hat{\mathbf{S}}]},} \\ 
& & S^{\mbox{\scriptsize \em eff}}[{\mathbf{N}}^{\bot}] \simeq
\int_0^{\infty} d\tau \left[ \frac{g}{2}\sum_{i \in A, \mu}
|{\mathbf{N}}^{\bot}_{i,\mu}|^2 \right. \nonumber \\
& & - \frac{g}{16}\left( \sum_{i \in A } |{\textstyle \sum_{\mu}}
{\mathbf{N}}^{\bot}_{i,\mu}|^2 + \sum_{j \in B} |{\textstyle
\sum_{\mu}} {\mathbf{N}}^{\bot}_{j-\mu,\mu}|^2 \right)
\nonumber \\  
& & \left. + \frac{1}{64g} \! \left( \sum_{i\in A} |{\textstyle \sum_{\mu}}
{\partial}_{\tau} {\mathbf{N}}^{\bot}_{i,\mu} |^2 + \sum_{j
\in B} |{\textstyle \sum_{\mu}} {\partial}_{\tau}
{\mathbf{N}}^{\bot}_{j-\mu,\mu} |^2 \! \right) \right],  
\end{eqnarray}
where $\tau$ is imaginary time and we have set the temperature to
zero.

Since we are interested in obtaining a long-wavelength theory for the
spin-waves, we make the coarse-graining: ${\mathbf{N}}^{\bot}_{i,\mu}
\rightarrow {\mathbf{N}}^{\bot}_i$, $i \in A$. This amounts to working
in the basis of the Goldstone modes of the theory. With this
approximation, we arrive at the effective action: 
\begin{eqnarray}
\lefteqn{ S^{\mbox{\scriptsize \em eff}}[{\mathbf{N}}^{\bot}]
\simeq \int_0^{\infty} d\tau   
\left[ \sum_i \left( \frac{3g}{4} |{\mathbf{N}}^{\bot}_i|^2 +
\frac{5}{16g} |{\partial}_{\tau} {\mathbf{N}}^{\bot}_i|^2
\right) \right. } \nonumber \\ 
&+& \sum_{\langle i,i' \rangle} \left( -\frac{g}{4}
{\mathbf{N}}^{\bot}_i \cdot {\mathbf{N}}^{\bot}_{i'} + \frac{1}{16g} 
{\partial}_{\tau} {\mathbf{N}}^{\bot}_i \cdot
{\partial}_{\tau} {\mathbf{N}}^{\bot}_{i'} \right) \nonumber \\ 
&+& \left. \sum_{\langle \langle i,i''\rangle \rangle} \left(
-\frac{g}{8} {\mathbf{N}}^{\bot}_i \cdot {\mathbf{N}}^{\bot}_{i''} +
\frac{1}{64g} {\partial}_{\tau} {\mathbf{N}}^{\bot}_i \cdot
{\partial}_{\tau} {\mathbf{N}}^{\bot}_{i''}\right) \right] ,   
\end{eqnarray}
where all sites are on the A sublattice, ${\scriptstyle \langle \cdots 
\rangle}$ refers to
nearest-neighbor pairs, and ${\scriptstyle \langle \langle \cdots
\rangle \rangle}$ refers to next-nearest-neighbor pairs.
Fourier transforming to $\mathbf{k}$ and (imaginary) $\omega$ gives:
\begin{equation}
S^{\mbox{\scriptsize \em eff}} = \int_{{\mathbf{k}},\omega}
g |{\mathbf{N}}^{\bot} ({\mathbf{k}},\omega)|^2 \left[ \left(
1-\gamma_{\mathbf{k}}^2 \right) +
\frac{\omega^2}{4g^2} \left( 1+ \gamma_{\mathbf{k}}^2\right) \right], 
\label{magnons}
\end{equation} 
with 
\begin{equation} 
\gamma_{\mathbf{k}} \equiv \frac{1}{2}(\cos k_x + \cos k_y) .
\label{mdispersioni}
\end{equation}  

This immediately gives the lowest-order magnon dispersion,
$\omega_{k} = 2g
\sqrt{(1-\gamma_k^2)/(1+\gamma_k^2)}$.
This dispersion is similar to the usual one found using, for instance,
Holstein-Primakov bosons (see, for instance, Ref.\cite{auerbach}), but
the ratio of the 
spin-wave velocity to the maximum of the dispersion is different. Even
neglecting spinons entirely, this is only a lowest-order result
for fluctuations of the field $\mathbf{N}$. Working to higher orders
in the perturbation theory of Eqns.(\ref{spqm1}-\ref{spqm2}) will
generate a more 
realistic spin-wave theory of the conventional antiferromagnet. For
the purposes of calculating the INS response, we satisfy ourselves
with the lowest-order result for the magnons. 

The lowest-order effect of
the spinons, as we have seen in Section \ref{scmft}, will be to reduce
$\langle {\mathrm{N}}^z \rangle$ from unity. At higher order, we
expect interactions with the spinons to further affect the magnon
dispersion. However, it would be wise for the purposes of comparing
the magnon and spinon INS responses to take into account this over-all
reduction of the N\'{e}el order by the
spinons. Therefore, we make the substitution: $g \rightarrow gN_0$,
where $N_0$ is set by the self-consistency condition,
Eq.(\ref{selfcon}), giving a ``self-consistent'' magnon dispersion:
\begin{equation}
\omega_{\mathbf{k}} = 2gN_0
\sqrt{\frac{1-\gamma_{\mathbf{k}}^2}{1+\gamma_{\mathbf{k}}^2}} \;. 
\label{mdispersionf}
\end{equation}

One important consequence of this calculation is that it allows one to
set the parameter $g$ in terms of experimentally-measured properties
of the magnon spectrum. Because it is probably less sensitive to
details of magnon interactions,
we choose to set $g$ by the maximum of the magnon dispersion rather
than by the spin-wave velocity near ${\mathbf{k}} = (0,0)$. In the
Heisenberg model, the maximum of the magnon dispersion is $2J$; we
therefore make contact with this model by identifying 
$J = gN_0$. Many experimental probes of the undoped cuprates find $J$
to be around 150 meV \cite{aeppli,ls&t,kastner,lorenzana}. 

\section{Inelastic Neutron Scattering Response}

What is the difference in spin
response between the model given in Eq.(\ref{separated}) and a
spin-charge confined antiferromagnet? The lowest energy spin
excitations in both systems are the spin-1 magnons dictated by
Goldstone's theorem in this symmetry-broken phase. For the confined
insulator, the lowest-energy spin-1/2 excitations would presumably be
something like single electrons, which have a huge Mott gap (on the
order of a few eV in the cuprates). In contrast, we see immediately that
provided one is in the regime where Eq.(\ref{separated}) holds
({\em i.e.} at temperatures and energies small compared to the vison
gap), the lowest energy spin-1/2
excitations are spinons which propagate as independent excitations
above the N\'{e}el (and \emph{d}-wave) gap which is on the order of $J
\sim $ 0.1 eV. In the previous sections, we
have laid out a theory of the spin degrees of freedom in a
fractionalized insulator with long-range N\'{e}el order. Now, we
calculate the INS signal in $AF^*$ using the lowest-order theories of the
magnons, Eq.(\ref{magnons}), and spinons, Eq.(\ref{MFT}).

\subsection{Magnetic Neutron Scattering Cross-Section}

The differential cross-section for neutrons scattering by wave-vector
${\mathbf{q}} = {\mathbf{k}}_f - {\mathbf{k}}_i$ 
and energy $\omega$  off electronic
spins is \cite{squires}:
\begin{equation} 
\frac{d^2\sigma}{d\Omega d\omega} \sim \frac{|k_f|}{|k_i|} F^2({\mathbf{q}}) 
\sum_{\alpha ,\beta }\left( \delta_{\alpha \beta} - \frac{q_{\alpha}
q_{\beta}}{q^2}\right) {\cal S}^{\alpha \beta}({\mathbf{q}},\omega) 
\end{equation} 
($\alpha $,$\beta = x,y,z$),
where $F^2({\mathbf{q}})$ is a form factor and the spin structure factor is: 
\begin{equation}
{\cal S}^{\alpha \beta }({\mathbf{q}},\omega ) = \frac{1}{\pi }
\frac{1}{1-e^{-\omega /k_B T}} \mbox{Im}{\chi }^{\alpha
\beta}({\mathbf{q}},\omega ), 
\end{equation}
where
\begin{equation}
\chi ^{\alpha \beta } ({\mathbf{q}},i{\omega }_{n} ) = \int_{0}^{\beta }
d\tau e^{i {\omega }_n \tau } \langle T_{\tau } {\hat{\mathrm{S}}}^{\alpha
}_{\mathbf{q}} (\tau) {\hat{\mathrm{S}}}^{\beta }_{-{\mathbf{q}}} (0) \rangle
\end{equation}
is the imaginary-time spin-spin
correlation function. At temperatures such that $k_B T \ll \omega$, we
can take the zero-temperature limit:
\begin{equation}
\frac{1}{1-e^{-\omega /k_b T}} \rightarrow \Theta(-\omega)
\end{equation}
with $\Theta (x)$ the Heavyside-step function. The spin-1/2
operators are given by the usual expression with electron operators
replaced by spinon operators: $\hat{\mathbf{S}}({\mathbf{q}},\tau) =
\sum_{\mathbf{k}} \hat{f}_{\mathbf{q}+\mathbf{k}}^{\dag}(\tau )
\mbox{\boldmath $\sigma$} \hat{f}_{\mathbf{k}}(\tau)$. 

The model outlined in the sections above provides a theory of the spin
response in $AF^*$. In the next section, we use the lowest-order
results to calculate the magnon and spinon signals (respectively)
in inelastic neutron scattering. We include a brief discussion of
higher-order effects.
  
\subsection{Magnon Response \label{magnon_signal}}

Starting from the $t_s = \Delta =0$ spin Hamiltonian, Eq.(\ref{HS}), it
is straightforward to calculate the spin-spin response function by
including a source term in the effective action of the form:
$\sum_i \hat{\mathbf{S}}_i \cdot {\mathbf{K}}_i$. The result, to
lowest order in fluctuations of $\mathbf{N}$ is: 
\begin{eqnarray}
{\cal S}_{\mbox{\scriptsize magnons}}^{\scriptscriptstyle
+-}({\mathbf{q}},\omega) &=& \frac{4}{(1+\gamma_{\mathbf{q}}^2)^2}  
\frac{(1- \gamma_{\mathbf{q}})^2}{\sqrt{1-\gamma_{\mathbf{q}}^4}}
\delta(\omega - \omega_{\mathbf{q}}),  
\label{magnonweight} 
\end{eqnarray}
with $\gamma_{\mathbf{q}}$ and $\omega_{\mathbf{q}}$ given in
Eqns.(\ref{mdispersioni}-\ref{mdispersionf}).  
While the exact form of this response function differs from that found
in, say, the Holstein-Primakov formalism, it has the same universal
features. By
universal properties we mean that the limits of both the response 
function and the spin-wave dispersion as ${\mathbf{q}} \rightarrow (0,0)$
and ${\mathbf{q}} \rightarrow (\pi,\pi)$ are the same for all
calculational methods because they are dictated by 
symmetries. Including intra-spin-wave interactions modifies the
non-universal aspects of the dispersion ({\em e.g.} the spin-wave velocity
near $\mathbf{q} = \mathbf{0}$) and yields direct multi-magnon
contributions to the 
spin structure factor. These direct multi-magnon processes are of a
much lower weight than the single magnon processes and so we ignore them.
 
To first order then, the magnon response to neutron scattering in 
$AF^*$ has the same universal features as in a conventional
antiferromagnet. This calculation allows us to fix the parameters in
our theory: we have seen in
Section \ref{fluctuations} that the energy scale $gN_0 = J$ to this
order. Additionally, the magnitude of the direct spinon signal can be
compared with the magnitude of this well-established magnon signal. 

\subsection{Spinon Response \label{spinon_signal}}

As detailed in Section \ref{mft}, at the mean field level, the spinon
part of the Hamiltonian in  
Eq.(\ref{separated}) is quadratic and has been solved elsewhere
\cite{arpes,gros}. 
The spin-flip and longitudinal structure factors for the spinons are
as follows:
\begin{eqnarray}
{\cal S}_{f}^{\scriptscriptstyle +-}({\mathbf{q}},\omega ) &=&
\int_{\mathbf{k}} \left[ \left( 1- \frac{\epsilon_{{\mathbf{q}} -
{\mathbf{k}}}}{{\cal E}_{{\mathbf{q}}-{\mathbf{k}}}} \right) \left( 1+
\frac{\epsilon_{\mathbf{k}}}{{\cal E}_{\mathbf{k}}} \right) -
\frac{\Delta_{{\mathbf{q}}-{\mathbf{k}}}\Delta_{\mathbf{k}}}{{\cal
E}_{{\mathbf{q}}-{\mathbf{k}}} {\cal E}_{\mathbf{k}}} \right. \nonumber \\ 
& & + \left. \frac{N_g^2}{{\cal E}_{{\mathbf{q}}-{\mathbf{k}}} {\cal
E}_{\mathbf{k}}} \right]  \times \delta (\omega - {\cal
E}_{{\mathbf{q}}-{\mathbf{k}}} - {\cal E}_{\mathbf{k}}), \\   
&=&{\cal S}_{f}^{\scriptscriptstyle -+}({\mathbf{q}},\omega ), \\
{\cal S}_{f}^{zz}({\mathbf{q}},\omega ) &=& \frac{1}{2}\int_{\mathbf{k}}
\left[ \left( 1- \frac{\epsilon_{{\mathbf{q}} - {\mathbf{k}}}}{{\cal
E}_{{\mathbf{q}}-{\mathbf{k}}}} \right) \left( 1+
\frac{\epsilon_{\mathbf{k}}}{{\cal E}_{\mathbf{k}}} \right) -
\frac{\Delta_{{\mathbf{q}}-{\mathbf{k}}}\Delta_{\mathbf{k}}}{{\cal
E}_{{\mathbf{q}}-{\mathbf{k}}} {\cal E}_{\mathbf{k}}}
\right. \nonumber \\ 
& & - \left. \frac{N_g^2}{{\cal
E}_{{\mathbf{q}}-{\mathbf{k}}} {\cal E}_{\mathbf{k}}} \right]  \times \delta
(\omega - {\cal E}_{{\mathbf{q}}-{\mathbf{k}}} - {\cal
E}_{\mathbf{k}}) \nonumber \\   
& & + \mbox{elastic (Bragg) response,} 
\label{oh good lord} 
\end{eqnarray} 
where $N_g, \epsilon_k, \Delta_k,$ and ${\cal E}_k$ are defined in Eqns.
(\ref{fdispersioni}-\ref{fdispersionf}).

These formulas have a few salient features.
Spin-flip neutron scattering leaves a spin-1 excitation in the
sample with momentum $\mathbf{q}$ and energy $\omega$. The
expression for ${\cal S}_{f}^{\scriptscriptstyle +-}$ simply sums the
ways of destroying a spin-down (-up) spinon at momentum $-\mathbf{k}$ and
creating a spin-up (-down) spinon at momentum $\mathbf{q}-\mathbf{k}$,
with the constraint 
that the energy cost for this process must be $\omega$. The rest of the
expression is the zero-temperature probability that the state at
$-\mathbf{k}$ is occupied and the one at $\mathbf{q}-\mathbf{k}$ is
unoccupied, appropriate for fermions. If one takes $N_g =0$
(corresponding to no N\'{e}el order), the full spin-rotation
invariance of the spinon system is restored and ${\cal
S}^{\scriptscriptstyle +-} = {\cal S}^{zz}$, as expected. 

For INS using unpolarized neutrons, the differential cross section is
a combination of these two signals, but it is possible using polarized
neutrons to obtain signals from these two channels separately (albeit
at a large cost to the intensity). It is worth noting that, provided one
aligns the polarized neutron spins along the direction of the
staggered magnetization, the signal due to single magnons is only in
the ``spin-flip'' sector, i.e. ${\cal S}_{\mbox{\tiny 1
magnon}}^{zz}=0$. The first contribution to  
${\cal S}^{zz}$ from spin-waves alone occurs in the two-magnon channel
which 
is of a greatly reduced weight compared to the single magnon signal.
The spinons then constitute 
the only magnetic signal of reasonable weight in this channel at
energy scales of order J. 

\begin{figure}
\vspace{0.2in}
\epsfxsize=2.8in
\centerline{\epsffile{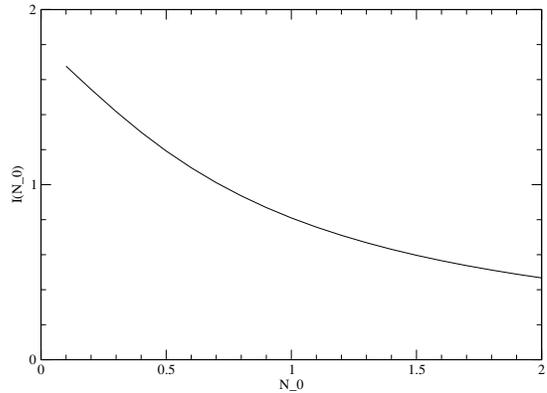}}
\vspace{0.21in}
\caption{Results of numerically integrating the RHS of
Eq.(\ref{selfcon}), called $I(N_0)$, for $t_s/g = \Delta/g = 0.7$. The  
self-consistent value of $N_0$ is the one for which $I=1$.}
\vspace{0.15in}
\label{selfconN0}
\end{figure}

A discussion of parameters is wise at this
point. Working in units of $gN_0 = J$, we will use $t_s=\Delta = J$ to
calculate ${\cal S}^{\alpha \beta }$ for the 
spinons. This gives numerical values of these constants which are 
reasonable for the undoped cuprates. However, one may wonder whether
they violate our above working assumptions $t_s/g, \Delta/g \ll
1$. Indeed, it is not immediately obvious how to set the parameter
$g$. Here, we take the following tack: the quantity 
$gN_0$ is set by comparison between the magnon dispersion in
Eq. (\ref{mdispersionf}) and the maximum of the magnon dispersion
in INS experiments. This gives $gN_0= J\simeq 150meV$. The parameters
$t_s$ and $\Delta$ may similarly be set by experiments and we use
the values given above. With these two constraints, the self-consistency
equation, Eq. (\ref{selfcon}), becomes an equation for the parameter
$g$. It is clear that if the values of $t_s$ and $\Delta$ are too
large, the equation for $g$ will have no solutions. For the parameters 
above,
using the expansion (valid for small $t_s/g$ and $\Delta/g$) in
Eq.(\ref{expansion}) as a
first iteration yields $g \simeq 1.4J$. Plugging the resultant values of
$t_s/g$ and $\Delta/g$ into the full self-consistent equation for
$N_0$, the right hand side of Eq.(\ref{selfcon}) can be numerically
integrated for various values of $N_0$. The result is shown in Fig.
\ref{selfconN0}. It is clear from the graph that the value of
$N_0$ which satisfies the self-consistent equation is $N_0 \simeq
0.7$. This gives $gN_0 = .98 J \simeq J$, and we have our
self-consistent parameters. We note that the value $N_0 \simeq 0.7$ is
still rather close to the ``no spinon'' mean field value of unity, so
that treating the spinons as a perturbation is somewhat justified
\cite{note}. 

The above expression for ${\cal S}_{f}^{\scriptscriptstyle +-}$ can be
numerically integrated to obtain the spinon response. This has been
performed, approximating the $\delta$-function in energy with a
Lorentzian of width $\epsilon = 0.01 J$. None of the salient features
of the results were influenced by the specific values of parameters
(such as $t_s, \Delta, \epsilon$).
 
\subsection{INS Signal \label{graphs}} 

\begin{figure}
\epsfxsize=3.5in
\centerline{\epsffile{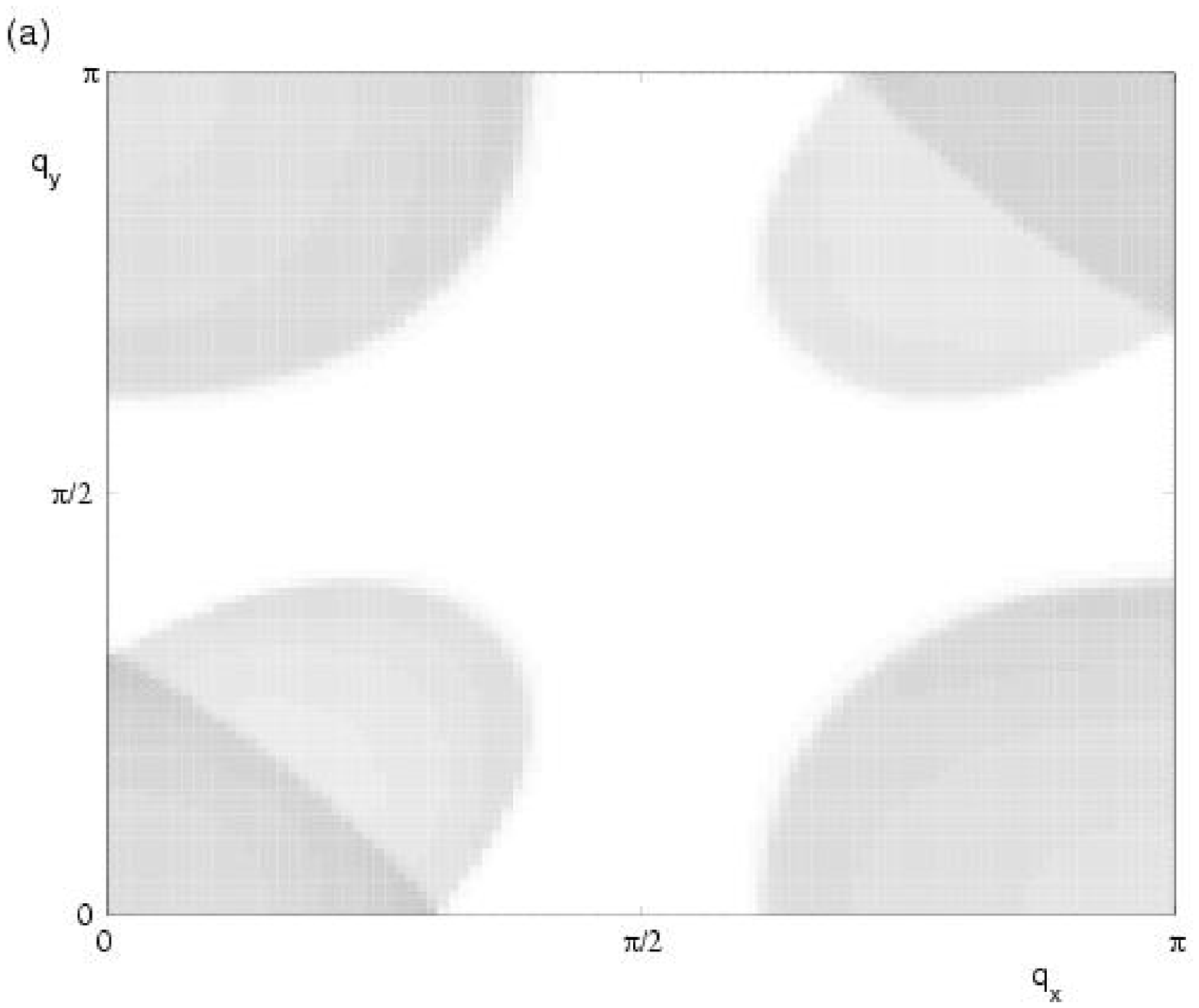}}
\vspace{0.10in}
\epsfxsize=3.5in
\centerline{\epsffile{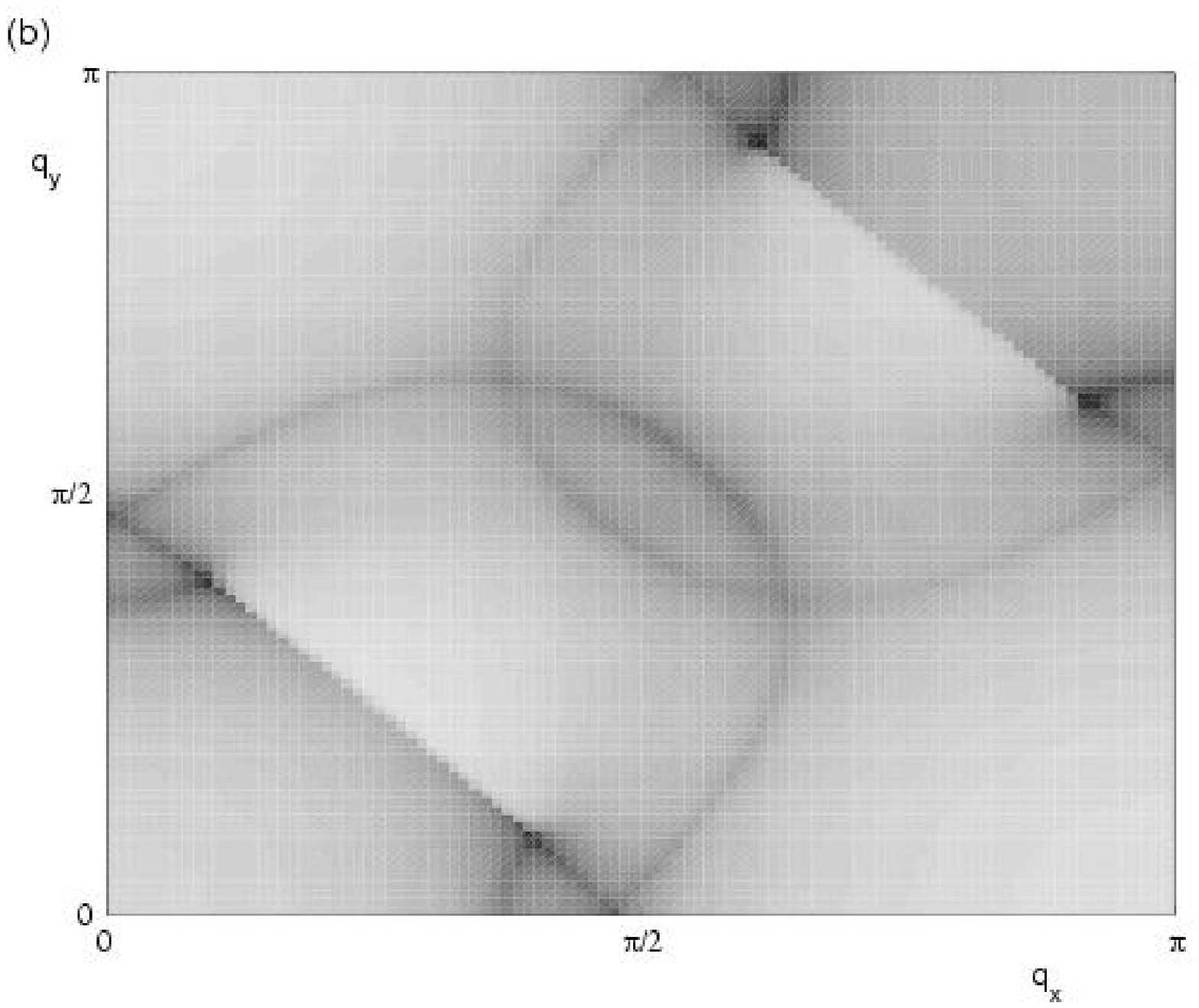}}
\vspace{0.10in}
\epsfxsize=3.5in
\centerline{\epsffile{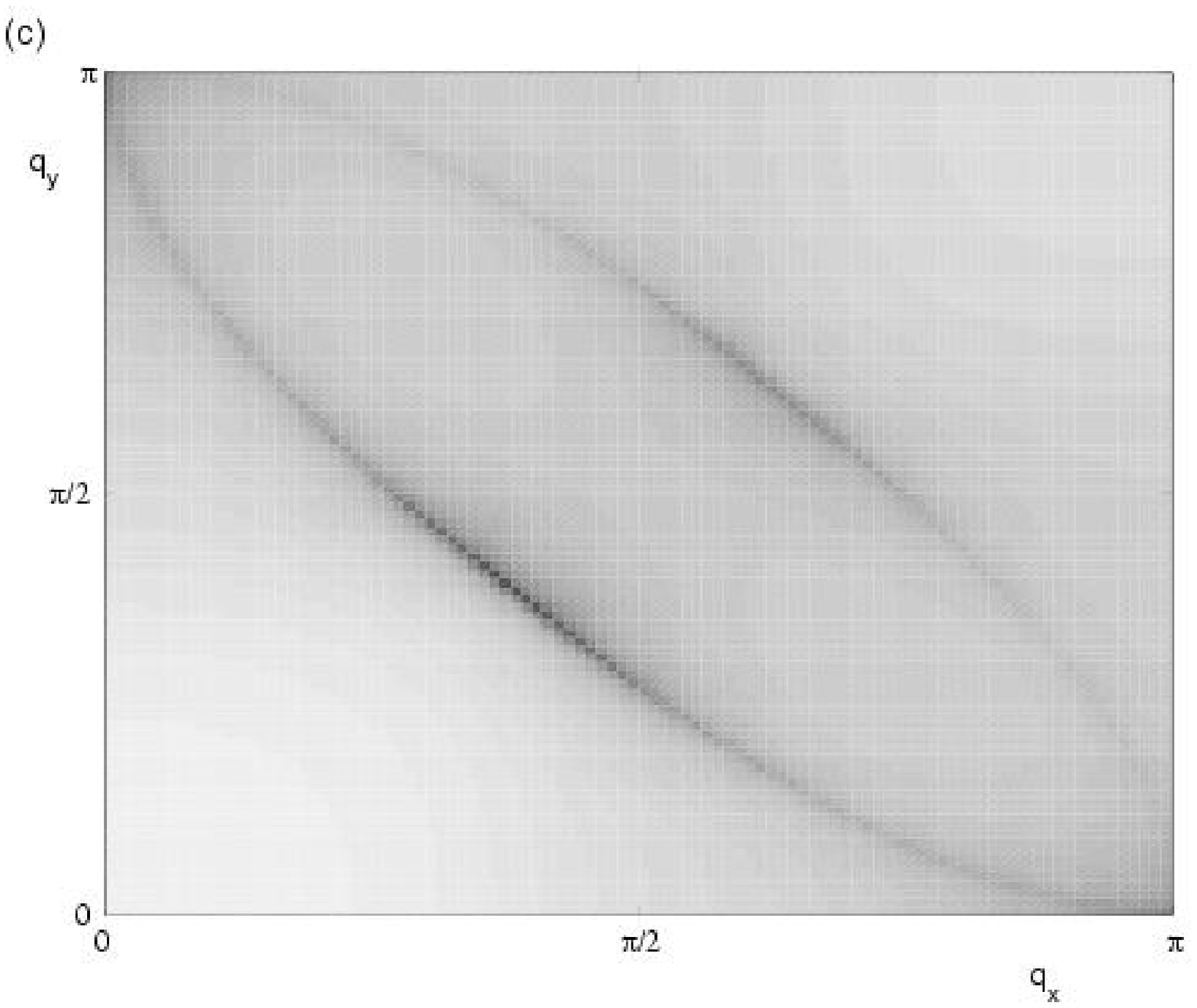}}
\vspace{0.10in}
\epsfxsize=3.5in
\centerline{\epsffile{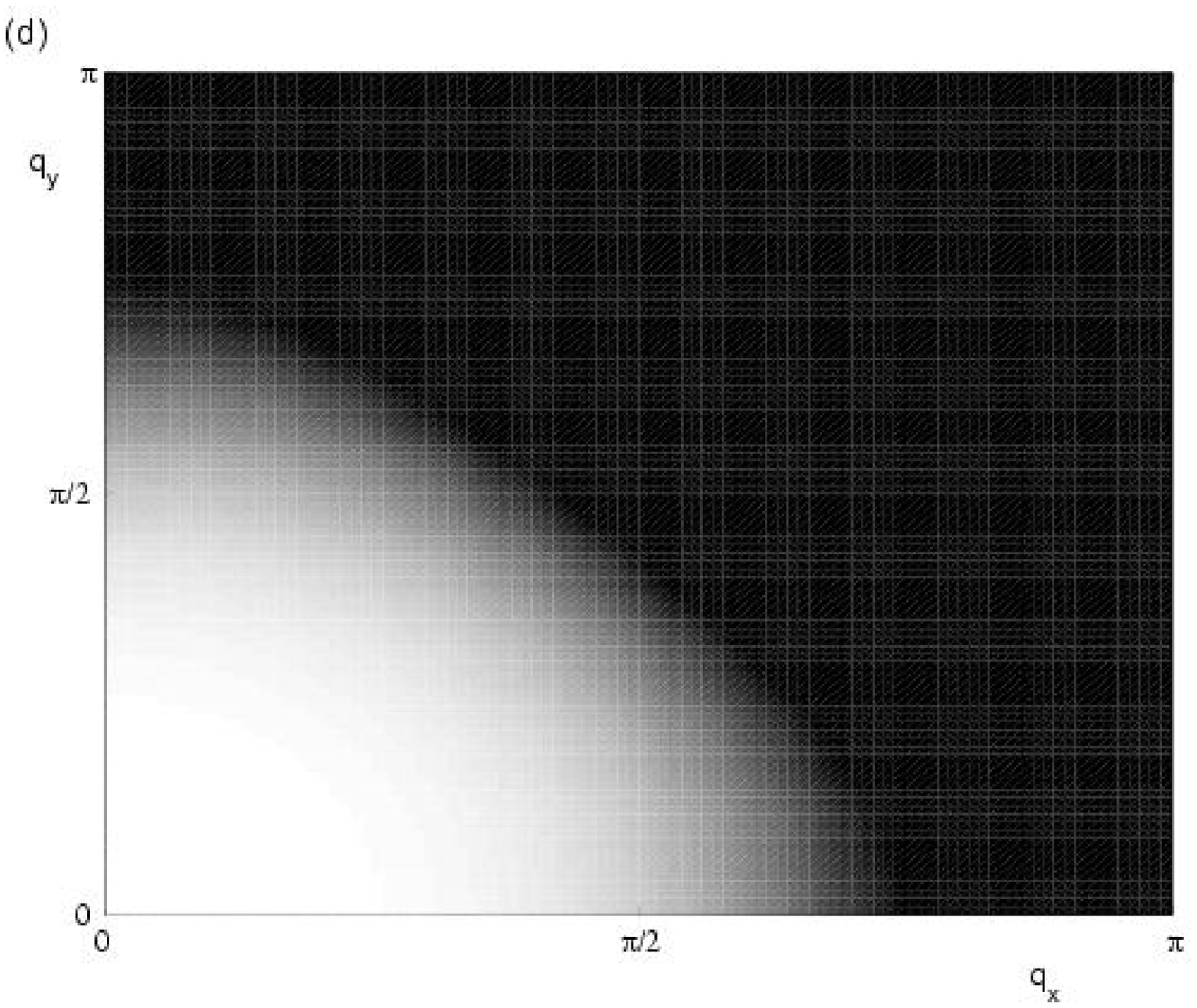}}
\vspace{0.15in}
\caption{Spinon ${\cal S}^{\scriptscriptstyle
+-}({\mathbf{q}},\omega)$ at energies $\omega$/J = (a) 4.5 (b)5.0
(c)6.0 . The single magnon weight, Eq.(\ref{magnonweight}), is shown
in (d). White regions are low intensity and black regions are high
intensity.}    
\vspace{0.15in}
\label{that} 
\end{figure}

In Figs. \ref{this} and \ref{that} we present the lowest-order
results for
${\cal S}^{\scriptscriptstyle +-} = {\cal S}^{\scriptscriptstyle
+-}_{f} + {\cal S}^{\scriptscriptstyle +-}_{\mbox{\tiny magnons}}$,
corresponding to spin-flip neutron scattering. All energies are in 
units of $J=gN_0$. 
Note that we have taken the staggered magnetization
to point in the $\hat{z}$-direction, but since our theory does not
contain terms which couple the spin and spatial variables (as, say,
a spin-orbit coupling would), the spin axes can be rotated independently
of the spatial axes.

\begin{figure}
\epsfxsize=2.5in
\centerline{\epsffile{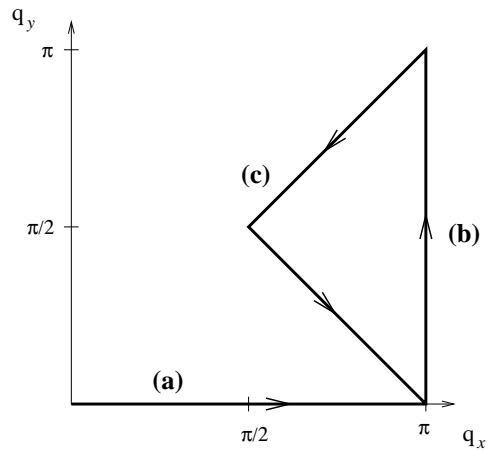}}
\vspace{0.15in}
\caption{Contours in the $q_x$, $q_y$ plane.}
\vspace{0.15in}
\label{cuts}
\end{figure}  

As ${\cal S}^{\alpha \beta}({\mathbf{q}},\omega)$ is a function of
three variables (for the 
effectively two-dimensional cuprates of interest), we show here a few
views of this function. Fig.\ref{this}, shows contour plots
of the intensity as a 
function of distance along cuts in the $(q_x,q_y)$ plane (shown in
Fig.\ref{cuts}) and energy. For the magnon signal,  
the $\delta$-function in Eq.(\ref{magnonweight}) has been replaced
with a ``box'' function:
\begin{equation}
\delta (x) = \left \{ \begin{array}{ll} 1/\epsilon & \mbox{for}
-\epsilon/2 < x < \epsilon/2 ,\\ 0 & \mbox{else}, \\ \end{array}
 \right. 
\end{equation}
with $\epsilon = 0.06 J \simeq 10meV$.
In these plots, we see the magnon dispersion with zeros at ${\mathbf{q}} =
(0,0)$ and $(\pi,\pi)$ and a vanishing weight near $(0,0)$. Above
this, we see the spinon continuum turn on with a lower-bound which is
modulated with twice the period of the magnon dispersion. 
We see in this plot that the spinon signal is
appreciable.  

\begin{figure}
\epsfxsize=3.5in
\centerline{\epsffile{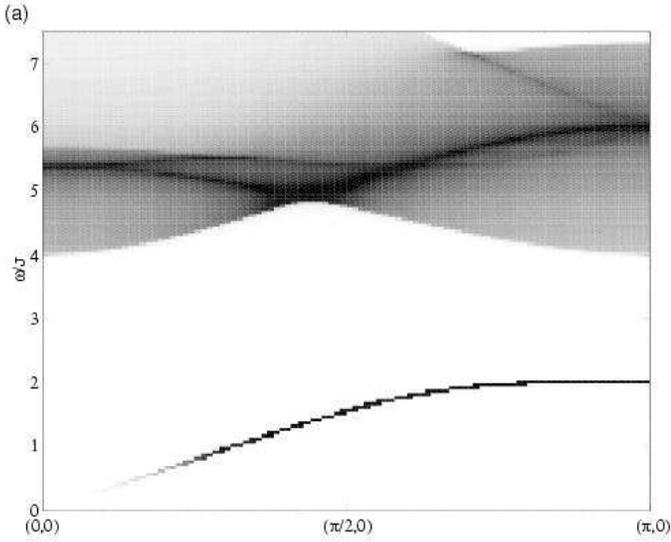}}
\vspace{0.10in}
\epsfxsize=3.5in
\centerline{\epsffile{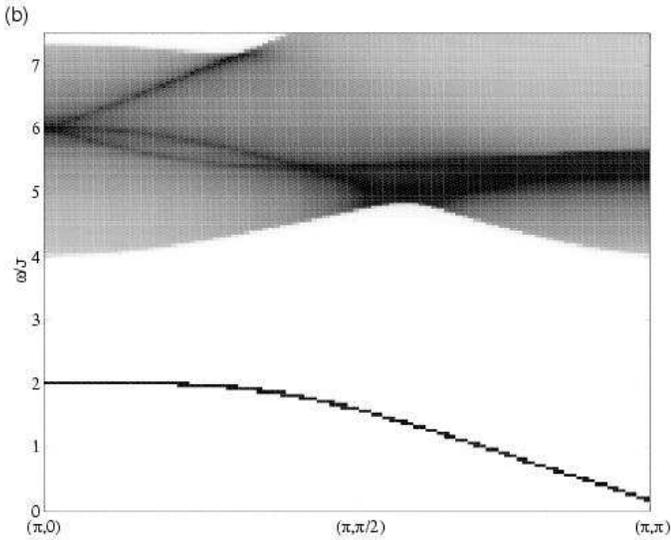}}
\vspace{0.10in}
\epsfxsize=3.5in
\centerline{\epsffile{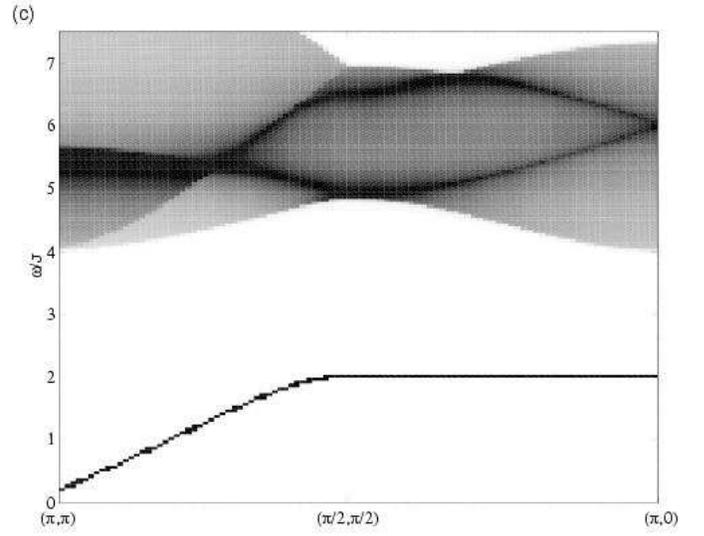}}
\vspace{0.15in}
\caption{${\cal S}^{\scriptscriptstyle +-}({\mathbf{q}},\omega)$ along cuts
(a),(b),and (c) in Fig. \ref{cuts}. Energies are in units of
$J$. White regions are zero intensity and black regions are high intensity.}
\vspace{0.15in}
\label{this}
\end{figure}  

In Fig.\ref{that}, we show
contour plots of the spinon intensity as a function of
$q_x$ and $q_y$ at constant values of the energy. We see that the
spinon intensity at turn-on ($\omega = 4J$) is largest near the
corners at $(0,0), (\pi,\pi)$, etc. For comparison,
Fig. \ref{that}(d) shows plots of the single magnon weight
(Eq.(\ref{magnonweight})) as a function of $q_x$ and $q_y$.

\begin{figure}
\epsfxsize=3.5in
\centerline{\epsffile{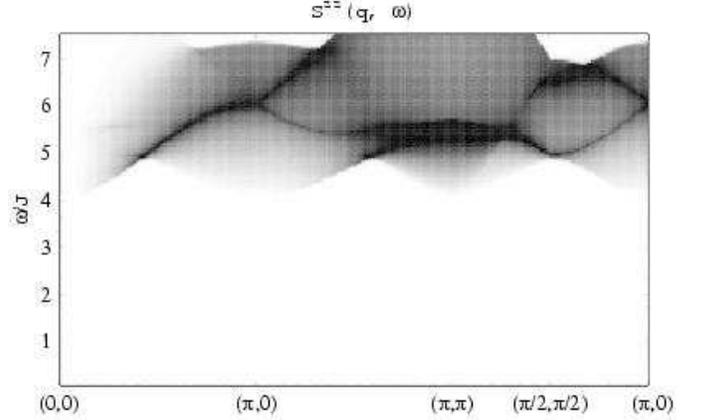}}
\vspace{0.15in}
\caption{${\cal S}^{\scriptscriptstyle zz}({\mathbf{q}},\omega)$ along
a cut (a)+(b)+(c) in Fig. \ref{cuts}. Energies are in units of $J$ and
the color scheme is the same as in Figs. \ref{this} and \ref{that}.}
\vspace{0.15in}
\label{zz} 
\end{figure}

In Fig. \ref{zz} we present contour plots of 
$S^{zz}_{f}({\mathbf{q}},\omega)$ sequentially along the cuts in Fig.
\ref{cuts}. We note again that, to first order, there is no magnon
contribution to this correlation function and so the two-spinon continuum
visible in these plots should in principle be the 
primary magnetic source of INS in this channel.

\subsection{Higher Order Effects}

At higher orders in the perturbation
theory of Eqns. (\ref{spqm1} -\ref{spqm2}), we would expect spin
wave interactions (present and important for detailed features even
when $t_s$ and $\Delta$ are zero) to modify the magnon dispersion and
lead to multi-magnon signals in both the ${\cal S}^{\scriptscriptstyle
+-}$ and ${\cal S}^{zz}$ channels. This calculation would be the same
as for a conventional antiferromagnet (since $t_s = \Delta =0$) and
would presumably lead to the actual spin-wave response in a
conventional antiferromagnetic system. At higher orders in $t_s/g$ and  
$\Delta /g$, we would obtain magnon interactions mediated by spinons,
which are not present in conventional antiferromagnets.
Being gapped excitations, the
spinons are inherently a ``high energy'' phenomena as far as the
spin-waves are concerned. We might therefore expect them to influence
most heavily the high-energy portions of the magnon dispersion. 
One can see from the graphs given in Section \ref{graphs} that the
lower edge of the spinon continuum has a minimum near $(\pi,0)$,
where the magnon dispersion has one of its maxima. One would expect
interactions between these two excitations to effect the dispersions
most heavily near these points of closest approach. 
While the effect of unconventional excitations such as spinons can in
principle be seen in the spin-wave response, we have chosen in this paper
to focus on the direct
spinon contribution to inelastic neutron scattering, shown in the
previous section.
The spin-waves will also affect the spinon signal at higher orders,
however, probably not in any dramatic fashion. 

\section{Conclusions}

In this paper we have presented a theory of the spin excitations of
the fractionalized antiferromagnet, $AF^*$, in the limit where the
visons can be ignored. This theory is Eq.(\ref{wholeshebang}).
This phase has both
long-ranged antiferromagnetic order and the topological order
associated with fractionalization \cite{topological}. It contains two
spin-carrying excitations: the spin-1 magnons and the spin-1/2
spinons, which interact with each other. The theory is well-defined and
can be solved in a controlled manner in the limit $t_s/g, \Delta/g \ll
1$. In this paper, we have found the lowest-order theories of these
two excitations and have calculated the dynamic spin-spin response
functions of each, appropriate for inelastic neutron scattering
experiments. 

To lowest order, the magnon signal is the same as in a conventional
antiferromagnet, but higher-order effects of the spinons should lead
to modifications of the dispersion, particularly near $(\pi,0)$. The
main anomalous feature of the INS signal from $AF^*$ is the presence
of a spinon continuum at energies $\sim 4J$, which exists even in the
longitudinal response, where the magnons are expected to be absent at
lowest-order.  

We are grateful to Gabe Aeppli, Radu Coldea, Steve Girvin, Steve
Nagler, Doug Scalapino, Alan Tennant, and Senthil Todadri for helpful
discussions.  
This research was supported by the NSF under Grants DMR-9704005 and
PHY-9907949.

\end{multicols}


\begin{references}
\bibitem{note} Indeed the values $t_s/g = \Delta/g = 0.7$, while not
$\ll 1$, are at least $<1$.
\end{references}
\end{document}